\documentclass{article}

\usepackage{microtype}
\usepackage{graphicx}
\usepackage{subfigure}
\usepackage{booktabs} 
\usepackage{tabularx}

\usepackage{hyperref}

\usepackage[accepted]{icml2022}

\icmltitlerunning{Self-supervision and Learnable STRFs for Age, Emotion, and Country Prediction}

\begin{document}

\twocolumn[
\icmltitle{Self-supervision and Learnable STRFs for Age, Emotion, and Country Prediction}



\icmlsetsymbol{equal}{*}

\begin{icmlauthorlist}
\icmlauthor{Roshan Sharma}{equal,ece}
\icmlauthor{Tyler Vuong}{equal,ece}
\icmlauthor{Hira Dhamyal}{lti}
\icmlauthor{Mark Lindsey}{ece}
\icmlauthor{Bhiksha Raj}{lti,ece}
\icmlauthor{Rita Singh}{lti}
\end{icmlauthorlist}

\icmlaffiliation{lti}{Language Technology Institute, Carnegie Mellon University, Pittsburgh, USA}
\icmlaffiliation{ece}{Department of Electrical and Computer Engineering, Carnegie Mellon University, Pittsburgh, USA}

\icmlcorrespondingauthor{Roshan Sharma}{roshansh@cmu.edu}
\icmlcorrespondingauthor{Tyler Vuong}{tvuong@andrew.cmu.edu}

\icmlkeywords{Speech Processing, Emotion Recognition, Multi-task Learning}

\vskip 0.3in
]



\printAffiliationsAndNotice{\icmlEqualContribution} 

\begin{abstract}
This work presents a multitask approach to the simultaneous estimation of age, country of origin, and emotion given vocal burst audio for the 2022 ICML Expressive Vocalizations Challenge \textsc{ExVo-MultiTask} track. The method of choice utilized a combination of spectro-temporal modulation and self-supervised features, followed by an encoder-decoder network organized in a multitask paradigm. We evaluate the complementarity between the tasks posed by examining independent task-specific and joint models, and explore the relative strengths of different feature sets. We also introduce a simple score fusion mechanism to leverage the complementarity of different feature sets for this task. 

We find that robust data preprocessing in conjunction with score fusion over spectro-temporal receptive field and HuBERT models achieved our best \textsc{ExVo-MultiTask} test score of 0.412.

\vspace{-.5cm}
\end{abstract}
\vspace{-.2cm}

\section{Introduction}
\vspace{-.2cm}

\label{intro}

The 2022 ICML Expressive Vocalizations Challenge Multitask High-dimensional Emotion, Age, and Country Task (\textsc{ExVo-MultiTask}) \cite{challenge} involves the simultaneous estimation of an individual's age, country of origin, and expressed emotion given only short recordings of vocal bursts from the individual.

Motivation for predicting these aspects of individuals from their voices comes in many forms. For example, prediction of age and country of origin from voice is of particular interest in forensic science and profiling circles, as a vocal signal is sometimes the only evidence left behind by a person of interest. Using machines to suggest further information from this evidence can be quite beneficial in these cases \cite{computational_forensics}.

As another example, estimation of age and emotion from an individual's voice may prove to be very useful for automated customer service phone lines, as elderly people may struggle to operate this type of technology, and people exhibiting high emotions may require a human customer service representative to handle their situation in an empathetic manner. In cases like either of these, automatic prediction of these crucial pieces of information about a caller can allow machines to manage complex situations more appropriately while still providing efficiency in regular situations \cite{phone_emotion}. Knowledge of an individual's country, age, and emotion information can be of similar use in a medical situation \cite{medical_emotion}.

Solutions to these individual tasks exist in the current literature, and continue to be actively researched in the field. The most recent approaches to these tasks have, not surprisingly, turned to deep neural networks. In the case of emotion recognition from speech, state of the art accuracy has been achieved by applying self-supervision to an upstream/downstream architecture \cite{ssemotion}. Age and country information, which is more closely related to traditional speaker identification, can be estimated by networks that utilize x-vector embeddings as input features \cite{xvectors}. It has been shown that these networks benefit greatly from strategic applications of multitask learning and transfer learning techniques \cite{joint_age_gender}.

Simultaneous solutions to these tasks, on the other hand, remain widely unexplored. While it is not logical to hypothesize that an individual's age, country of origin, and choice of emotion are dependent or correlated variables, it may be the case that age or country of origin have an effect on the way an individual expresses a certain emotion. Because of this possible conditional relationship, it follows that performing all three of these tasks simultaneously in a multitask framework may be advantageous. This is the hypothesis which this work attempts to prove.

\section{Data Analysis and Preprocessing}
\vspace{-.2cm}

\label{data}
Data for the \textsc{ExVo-Multitask} track comes from the Hume Vocal Burst Competition Dataset (H-VB) \cite{Cowen2022HumeVB}. The dataset is composed of 59,201 samples of vocal burst audio, split nearly evenly into training, validation, and test sets. Emotion, age, and country labels are provided for the training and validation sets.

Audio data for the task was provided in two forms. First, contestants were given the preprocessed data which was used to train the baseline systems. This data was in \texttt{.wav} format with a sample rate of 16 kHz. Note that this data exhibited considerable clipping in the regions that contained vocalization, likely as a result of the preprocessing. As such, the original raw data was provided in the form of compressed \texttt{.webm} files in addition to the preprocessed data.

In this work, the following preprocessing steps were applied to the raw data: First, each sample was converted from the \texttt{.webm} filetype to \texttt{.wav} and downsampled to 16 kHz via \texttt{ffmpeg}. Second, the DC component of each sample was removed by applying a highpass finite impulse response (FIR) filter with a stopband of 20 Hz, after which each sample was gain normalized. Third, the noise in the audio was suppressed by applying a minimum-mean squared error short-time spectral amplitude (MMSE-STSA) filtering algorithm \cite{MMSE-STSA}, which is described in some detail in the following subsection. Finally, silence before and after the vocal bursts was removed using a voice activity detector (VAD) \cite{SileroVAD}.
\vspace{-.2cm}

\subsection{Denoising Using the Ephraim-Malah Algorithm}
\vspace{-.2cm}

Ephraim and Malah's MMSE-STSA filtering algorithm is based the principle that perceptual quality is more closely tied to amplitude than phase. As such, the spectral amplitude is optimized via MMSE and combined with the noisy phase to reconstruct an enhanced signal.

The MMSE-STSA is derived starting with the expression for the expected value of the clean STSA given the noisy short-time Fourier transform (STFT). After a series of approximations and algebraic manipulations, the expression in Eq. \ref{eq:mmse} is achieved for high signal-to-noise ratios (SNR):
\begin{equation} \label{eq:mmse}
    \hat{A}_k \cong \frac{\xi_k}{1+\xi_k}R_k
\end{equation}
Here, $\hat{A}_k$ and $R_k$ are the estimated clean spectral amplitude and given noisy spectral amplitude at time frame $k$, respectively, and $\xi_k$ is the SNR at frame $k$. From this expression, it is evident that when SNR is high, the multiplicative gain applied to the noisy spectral amplitude resembles a Wiener filter. However, when SNR is low, the gain deviates from the Wiener filter, avoiding the high bias of the Wiener filter at low SNRs.

The MMSE-STSA algorithm was used here because of its versatility in handling different SNRs. It was also shown to leave colorless residual noise rather than musical noise, which may be advantageous for the downstream classifier.
\vspace{-.2cm}

\subsection{Comparing Preprocessing Techniques}
\vspace{-.2cm}

To understand which preprocessing techniques would work best for the task at hand, we tested the data at multiple preprocessing stages. Specifically, we used the clipped baseline audio, un-clipped raw audio, DC normalized audio, denoised audio, and VAD-processed audio. The model architecture used here is the one provided by the challenge.

Table \ref{tab:compare16} shows the results achieved when openSMILE ComParE 2016 features are extracted from each version of the data. Scores are shown in terms of the evaluation metrics outlined for the \textsc{ExVo-MultiTask} track, including unweighted average recall (UAR) for country prediction, mean absolute error (MAE) for age estimation, concordance correlation coefficient (CCC) for emotion prediction, and the harmonic mean of these metrics ($S_{MTL}$). From the table we can observe that denoising the data is indeed important for the task and achieves good results. VAD degrades the performance, indicating that the silences in between the expressions actually carry useful information which, when removed, reduce the predictive capacity of the models.

\begin{table}[]
    \footnotesize 
    \centering
    \begin{tabular}{l|r|r|r|r}
    \hline
       Data & \multicolumn{1}{l|}{\begin{tabular}[c]{@{}l@{}}Country\\ (UAR)\end{tabular}} & \multicolumn{1}{l|}{\begin{tabular}[c]{@{}l@{}}Age\\ (MAE)\end{tabular}} & \multicolumn{1}{l|}{\begin{tabular}[c]{@{}l@{}}Emotion\\ (CCC)\end{tabular}} & \multicolumn{1}{l}{\begin{tabular}[c]{@{}l@{}}Val \\ $S_{MTL}$\end{tabular}} \\ 
       \hline
       Baseline & 0.416	& 0.506 & 4.22 & 0.348 \\
       Raw  & 0.443 & 0.510 & 4.30 & 0.352 \\
       DC Normalized & 0.514 & 0.433 &	4.47 &	0.344 \\
       Denoised & 0.515 & 0.449 & 4.27 & \textbf{0.355} \\
       VAD & 0.467 & 0.391 & 4.56 & 0.324 \\
       \hline
    \end{tabular}
    \vspace{-.2cm}

    \caption{A comparison of validation results when different preprocessing techniques are applied. We extract ComParE 2016 features from each dataset and train the multitask model architecture provided by the organizers. We find that denoising gives better results.}
    \label{tab:compare16}
    \vspace{-.4cm}

\end{table}
\vspace{-.2cm}
\section{Proposed Method}
\vspace{-.2cm}

Our sequence architecture is comprised of a feature frontend, encoder, and pooling multi-task decoder. 
\vspace{-.2cm}

\subsection{Feature Frontends}
\vspace{-.2cm}

Although the official challenge provided a few baseline utterance-level feature frontends, we wanted to explore the benefits of sequence-level feature frontends.  We decided to use standard log-mel spectrogram feature frontend, a modulation-based feature frontend and a HuBERT-based feature frontend.  The three feature frontends are described in further detail below.  
\vspace{-.2cm}
\subsubsection{Log-mel spectrogram feature}
\vspace{-.2cm}

The first sequence-level feature we tried was the standard log-mel spectrogram.  We chose the log-mel spectrogram since it is one of the most commonly used features for deep-learning-based speech processing and has been proven useful for many speech-related tasks.  In this study, the log-mel spectrogram is calculated using a 25-ms Hanning window with a 10-ms hop between frames and a 512-point discrete Fourier transform. 80 mel channels were calculated and we apply power compression using the logarithm subsequently.  Additionally, global mean and variance normalization is applied to each of the 80 mel channels.  
\vspace{-.2cm}

\subsubsection{Spectro-temporal modulation feature}
\vspace{-.2cm}

In addition to the standard log-mel spectrogram, we tried using a frontend feature that captures  spectro-temporal modulations.  Specifically, we use spectro-temporal receptive fields (STRFs) \cite{chiMultiresolutionSpectrotemporalAnalysis2005,meyerRobustnessSpectrotemporalFeatures2011} to capture the spectro-temporal modulations in the speech signal.  STRFs are believed to respond to a range of temporal and spectral modulation patterns in the auditory system.  Each STRF is parameterized  by a rate and scale parameter that control the temporal and spectral modulation selectivity, respectively.  Example of gabor-based STRFs are shown below in Figure \ref{fig:strf-kernels}.  
Inspired by the results in \cite{xia2021temporal} that used  STRFs to capture spectro-temporal modulations for emotion recognition, we developed a similar spectro-temporal modulation feature for multi-task learning.  We describe the spectro-temporal modulation feature below.   

Given $N$ STRFs, we define the spectro-temporal modulation feature (STMF) at time
frame $t$ and frequency bin $k$ to be
\begin{equation}
     \textrm{STMF}[i,t,k] = \textrm{STRF}^{(i)}[t,k] \star \textrm{LM}[t,k],
 \label{eq:STM}
\end{equation}
where $\textrm{LM}$ denotes the log-mel spectrogram, $\star$ refers to cross-correlation, and $i$ is the STRF index. Each of the $N$ STRFs are tuned to specific spectro-temporal modulation patterns.  We follow  \cite{vuong2020learnable} and enable the STRF modulation parameters to be a learnable parameter by the deep neural network (DNN).  Specifically, the STRFs are implemented as a 2D convolutional layer where the  DNN learns the rate and scale value that parameterize the STRF rather than the individual kernel weights.  The gabor-based STRFs are defined as the product between a 2D complex exponential and a hanning envelope. Having the STRF parameters learnable by the DNN enables us to learn the optimal modulation parameters directly without having to choose the optimal parameters beforehand.  The resulting STRF feature is 3D where the first dimension represents each of the spectro-temporal modulations  and the remaining two represent the time and frequency, respectively.

\begin{figure}[htb]
\centering
\includegraphics[scale=.15]{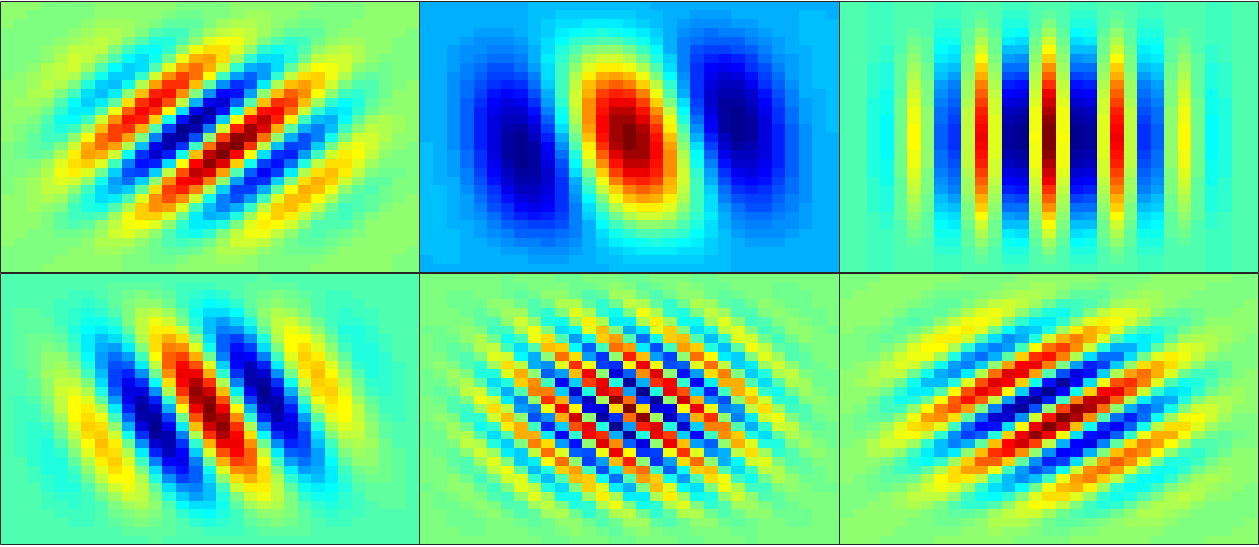}
\caption{Examples of Gabor-based STRF kernels tuned to various spectro-temporal modulation frequencies.}
\label{fig:strf-kernels}
\vspace{-.2cm}

\end{figure}




\vspace{-.2cm}
\subsection{Sequence Model Architecture}
\vspace{-.2cm}

\textbf{HuBERT Encoder}: The HuBERT \cite{Hsu2021HuBERTSS} encoder comprises a series of BERT-like transformer layers following a \texttt{wav2vec2.0} feature extractor unit. 

\textbf{Conformer Encoder}: Conformers are the state of the art on multiple speech tasks including speech recognition. Conformers comprise transformer like layers with macaron-style feed-forward layers, convolution, and self-attention.  The initial layer of the conformer encoder consists of a 2D convolutional downsampling layer.  We increase the input channels of the downsampling layer when STRF features are used.  

\textbf{Self-Attentive Pooling Decoder}: Our decoder uses task-specific self-attentive pooling, followed by a series of fully-connected layers. 

 \vspace{-.2cm}
\subsection{Training setup}
\vspace{-.2cm}

 We use the mean absolute error (MAE) objective function to compute the loss for both the emotion and age prediction.  For the country prediction, we use the negative log-likelihood objective function.  The final loss used to optimize our multitask system is a weighted combination of the three loss terms.  We tune the weights for each loss term based on the performance on the validation set.   \begin{equation}
     L_{total} = \alpha_1 L_{age} + \alpha_2 L_{emo} + \alpha_3 L_{country}
      \end{equation}

\vspace{-.3cm}

Due to each loss having a different range, we use 1, 80, and 8 for $\alpha_1$, $\alpha_2$, and $\alpha_3$, respectively.

This research was conducted using the ESPnet framework \cite{watanabe2018espnet}.  The code for the features and model architecture will be accessible here\footnote{https://github.com/espnet/espnet}.
\vspace{-.2cm}
\section{Experimental Results and Analysis}
\vspace{-.2cm}

\subsection{Joint Versus Independent Modeling}
\vspace{-.2cm}

\begin{table}[t]
\begin{tabular}{l|r|r|r}
\hline
Model Description & \begin{tabular}[c]{@{}l@{}}Country\\ (UAR)\end{tabular} & \multicolumn{1}{l|}{\begin{tabular}[c]{@{}l@{}}Age\\ (MAE)\end{tabular}} & \multicolumn{1}{l}{\begin{tabular}[c]{@{}l@{}}Emotion\\ (CCC)\end{tabular}} \\ \hline
Emotion-only & $\times$ & $\times$ & \textbf{0.626} \\
Age-only & $\times$ & 4.02 & $\times$ \\
Country-only & 0.588 & $\times$ & $\times$ \\ 
\hline
Joint Multitask & \textbf{0.626} & \textbf{3.97} & 0.622 \\ \hline
\end{tabular}
\caption{Performance of independent and multitask HuBERT encoder models on the \textsc{ExVo-MultiTask} validation set}
\label{tab:joint_results}
\vspace{-.2cm}

\end{table}

We hypothesize that the inter-relatedness between the tasks under consideration make multitask learning a viable option. 
We compared the use of task independent models and task dependent models to establish if this holds for the task. Table \ref{tab:joint_results} reports results on the validation set for our models that use the HuBERT encoder for each of the three tasks, and a simple multitask setup. 

We find that multitask learning improves performance on age regression and country classification, while retaining comparable emotion recognition performance. From these results, it appears that there is no clear dependence on country of origin and age for emotion prediction. This is likely to be more true for vocal bursts that do not contain spoken language, since expressive vocal bursts are likely similar across countries and age groups. 

\vspace{-.2cm}
\subsection{Individual Feature Results}
\vspace{-.2cm}

Table \ref{tab:ind_feats_result} summarizes the results on the validation set for each of the individual features, encoders and frontends.  For the log-mel and STRF features, we only use the original unprocessed waveform provided by the challenge.  For the HuBERT features, we use the original unprocessed waveform and also investigate the benefits of using denoised speech. On the validation set, all three features outperformed the challenge baseline.  Interestingly, we also observed additional benefits when we used the HuBERT encoder with the denoised speech.  We hypothesize that denoising the speech is beneficial since background noise decreases the intelligibility of speech.  Furthermore, we investigated the impact of speed perturbations for this task and found that it slightly improves country prediction and emotion recognition while performing worse on age prediction. 

In addition to analyzing the performance of the individual features, we explore whether the features provide complementary benefits.  Since each of the features extract information from different domains, we believe there will be benefits from combining different feature scores.  We discuss our score fusion results in the next section.  

\begin{table}[t]
\begin{tabular}{l|r|r|r|r}
\hline
Model & \multicolumn{1}{l|}{\begin{tabular}[c]{@{}l@{}}Country\\ (UAR)\end{tabular}} & \multicolumn{1}{l|}{\begin{tabular}[c]{@{}l@{}}Age\\ (MAE)\end{tabular}} & \multicolumn{1}{l|}{\begin{tabular}[c]{@{}l@{}}Emotion\\ (CCC)\end{tabular}} & \multicolumn{1}{l}{\begin{tabular}[c]{@{}l@{}}Val \\ $S_{MTL}$\end{tabular}} \\ 
\hline
HuBERT & 0.641 & 4.01 & 0.552  & 0.406 \\
~+ Denoise & 0.648 & \textbf{3.76} & 0.546 & \textbf{0.420} \\
~+ sp. perturb & \textbf{0.650} & 3.82 & \textbf{0.559} & 0.419 \\
\begin{tabular}[c]{@{}l@{}}Conformer \\ ~(STRF)\end{tabular} & 0.504 & 4.03 & 0.507 & 0.375 \\
\begin{tabular}[c]{@{}l@{}}Conformer \\ ~(log-mel)\end{tabular} & 0.496 & 4.17 & 0.523 & 0.370 \\
\hline
\end{tabular}
\caption{Performance of multitask models on the validation set for various encoders and feature frontends}
\label{tab:ind_feats_result}
\vspace{-.4cm}

\end{table}
\vspace{-.2cm}

\subsection{Score Fusion Results}
\vspace{-.2cm}

Table \ref{tab:score_fusion_result} summarizes the results on the test set for the fused scores from the individual features.  For score fusion, we simply take a weighted combination of the predictions from the individual features.  We used the weighted combination of the individual feature scores that provided the best results on the validation set.  Interestingly, when we combine the HuBERT predictions with either the log-mel conformer  or STRF conformer predictions, simply averaging the country and age predictions between the two gave us the best results on the validation set. For the emotion task, using the HuBERT emotion predictions without any fusion gave us the best results on the validation set. In our initial score fusion experiments, we found that fusing the HuBERT predictions with the STRF predictions improved the performance of the challenge score from .406 to .42 on the validation set. Although simple score fusion gave us improvement on the validation set, we found that the benefits on the test set were minimal.  We believe that training an additional classifier to combine the scores could provide benefits on the test set.  Additionally, we hypothesize more sophisticated deep learning-based fusion methods would be beneficial and generalize better on the test set.  

\begin{table}[]
\vspace{-.2cm}

\begin{tabular}{l|r|r|r|r}
\hline
Models & \multicolumn{1}{l|}{\begin{tabular}[c]{@{}l@{}}Country\\ (UAR)\end{tabular}} & \multicolumn{1}{l|}{\begin{tabular}[c]{@{}l@{}}Age\\ (MAE)\end{tabular}} & \multicolumn{1}{l|}{\begin{tabular}[c]{@{}l@{}}Emotion\\ (CCC)\end{tabular}} & \multicolumn{1}{l}{\begin{tabular}[c]{@{}l@{}}Test \\ $S_{MTL}$\end{tabular}} \\ \hline
HuBERT & 0.659 & 4.00 & 0.559 & 0.410 \\
\begin{tabular}[c]{@{}l@{}}Conformer \\ ~(STRF)\end{tabular} & 0.518 & 4.34 & 0.475 & 0.358 \\
\begin{tabular}[c]{@{}l@{}}HuBERT \\ + STRF\end{tabular} & 0.650 & 3.94 & 0.556 & 0.412 \\
\hline
\end{tabular}
\caption{Performance of multitask models on the denoised test set using score fusion between the STRF and HuBERT models}
\label{tab:score_fusion_result}
\end{table}

\vspace{-.2cm}
\subsection{Analysis of Best Model Result}
\vspace{-.2cm}

From our best test score, we observe that the model obtains the highest CCC on awe, amusement, and surprise. This is likely because of how distinctive the expression of these emotions is, while emotions such as awkwardness and triumph are more confusable. 

\vspace{-.2cm}
\section{Conclusion}
\vspace{-.2cm}

In this paper, we examined the task of jointly predicting emotional state, speaker age, and country of origin from expressive vocal bursts for the \textsc{ExVo-MultiTask} track. 
The use of sequence-based features outperforms utterance-level functionals reported in the baseline, which highlights the importance of time sequence information for the three tasks under consideration. We evaluate the relative strengths of different features---log-mel, STRFs, and content-based self-supervised HuBERT features---for each of these tasks and find that while learnable STRF kernels outperform log-mel features, HuBERT features exhibit superior performance. 

To exploit complementary across feature sets, we leverage a simple weighted combination-based score fusion of predictions of different models, thereby achieving our best performance with STRF and HuBERT features. In the future, we plan on exploring more sophisticated deep learning-based fusion methods.  Specifically, we plan on directly fusing the multiple input features as well as fusing intermediate output layers between multiple systems.

\newpage






\bibliographystyle{icml2022}
\bibliography{main.bib}

\end{document}